\documentclass[prl,showpacs,floatfix,twocolumn,nofootinbib]{revtex4}
\usepackage{graphicx}
\usepackage{bm,amsmath}
\usepackage{dcolumn}
\usepackage{natbib}
\begin{document}

\title{Proposal for a sensitive
search for electric dipole moment of electron with matrix-isolated radicals}

\author{M. G. Kozlov}
\affiliation{Petersburg Nuclear Physics Institute, Gatchina
188300, Russia}
\author{Andrei Derevianko}
\affiliation{Department of Physics, University of Nevada, Reno,
Nevada 89557, USA}

\begin{abstract}
We propose using matrix-isolated paramagnetic diatomic molecules to
search for the electric dipole moment of electron (eEDM). As was
suggested by Shapiro in 1968, the eEDM leads to a magnetization of
a sample in the external electric field.  In a typical condensed matter experiment,
the effective field  on the unpaired electron  is of the same order of magnitude as the
laboratory field, typically about $10^{5}$V/cm. We exploit
the fact that  the effective electric
field inside  heavy polar molecules is  in the order of $10^{10}$V/cm.
This leads to a huge enhancement of the Shapiro effect. Statistical
sensitivity of the proposed experiment may allow one to improve the current
limit on eEDM by three orders of magnitude in few hours
accumulation time.
%
%
\end{abstract}


\maketitle


The searches for the elusive
 electric dipole moment of electron (eEDM)
  are motivated by the fact that
the existence of a permanent EDM of a particle violates both parity
(P) and time-reversal (T) symmetries. Due to the compelling
arguments of the CPT theorem, the T-violation implies CP-violation,
a subject of great interest in the physics of fundamental
interactions~\cite{BigSan00}.
Current experimental limit on
eEDM~\cite{RCS02} is close to the predictions of many extensions to
the Standard Model of elementary particles,
such as``naive'' supersymmetry (SUSY)~\cite{ForPatBar03,Jun05}.
Other SUSY extensions yield eEDM a few orders of magnitude below the
present limit. Here we propose an eEDM search that may constrain
eEDM at that important level. Our proposal relies on the fact that
the thermodynamically averaged eEDM (and thus the electron's
magnetic moment aligned with eEDM) is oriented along the electric
field. We propose to employ polarized molecular radicals frozen in a
rare-gas matrix and measure the eEDM-induced magnetic field
generated by the sample. Conservative estimates project that the
present limit on eEDM can be improved by several orders of
magnitude.

The present limit on eEDM,
\begin{equation}
|d_e| < 1.6 \times 10^{-27} e \cdot \mathrm{cm} \,, \label{Eq:deTl}
\end{equation}
is derived from a high-precision measurement \cite{RCS02} with a
beam of Tl atoms. In such experiments one spectroscopically searches
for a tiny eEDM-induced splitting of the magnetic sublevels of an
atom in an externally applied electric field.

\begin{figure}[ht]
\begin{center}
\includegraphics[scale=0.5]{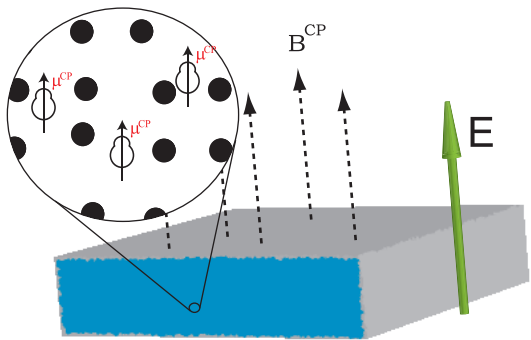}
\end{center}
\caption{\label{Fig:setup} Scheme of searching for EDM of electron
with diatomic radicals embedded in a matrix of rare-gas atoms. A
polarizing electric field $E$ is applied to the matrix. As a result,
molecular CP-violating magnetic moments $\mu^\mathrm{CP}$ become
oriented and generate ultraweak magnetic field $B^\mathrm{CP}$. By
measuring $B^\mathrm{CP}$ one places constraints on eEDM. }
\end{figure}

New atomic eEDM experiments plan to use optical trapping
\cite{CLV01}. There are two other major trends aimed at improving
the experimental sensitivity to eEDM: (i) employing molecules
instead of atoms in spectroscopic experiments
\cite{HST02,Hud05,DBB00} and (ii) non-spectroscopic solid state
experiments \cite{Lam02,MDS03,Hei05}. Here we propose to merge these
two trends by searching for eEDM with molecules trapped in a cold
matrix of rare-gas atoms (see Fig.~\ref{Fig:setup}). We argue that
this scheme combines advantages of both techniques. Indeed, the eEDM
effects in molecules are markedly amplified because of the strong
internal molecular electric field \cite{SF78}, much larger than
attainable laboratory fields. In the present solid-state schemes the
atomic enhancement of the external electric field for ions of a
solid is of the order of unity \cite{MDS03}. By using
matrix-isolated diatomic radicals, one can gain up to six orders of
magnitude in the effective electric field. At the same time one
retains a great statistical sensitivity of the solid-state searches.
We show that this particular combination seems to  drastically
improve sensitivity of the eEDM search.

Let us review important aspects of the non-spectroscopic solid-state
search for the eEDM. Introduced by \citet{Sha68}, this scheme
exploits the link between EDM of the electron and it's spin, $\bm{d}
= d_e \bm{\sigma}$, and therefore it's magnetic moment, $\bm{\mu}_e
\approx  -\mu_B \bm{\sigma} = -\mu_B\bm{d}/d_e$. In an external
E-field, because of the coupling of the eEDM to the E-field, thermal
populations of the spin-up and spin-down states slightly differ,
leading to the magnetization of the sample. By measuring the
generated magnetic field one derives constraints on the eEDM. A
proof-of-concept experiment has been carried out in 1978 by
\citet{VK78}. At that time the solid-state experiment appeared to be
less sensitive to the eEDM than the spectroscopic beam experiments.
It is only very recently that the advances in magnetometery (see
\cite{KomKorAll03} and references therein) have revived an interest
to the solid-state eEDM searches \cite{Lam02,Bar04,MDS03,DK05}.
Alternatively, one can look for a voltage induced in a sample in
external magnetic field~\cite{Hei05}.

We focus on molecular radicals (i.e., molecules with unpaired spin)
in the ground $^2\Sigma_{1/2}$ state. Consider a sample of radicals
in thermodynamic equilibrium at temperature $T$. Because of the eEDM
coupling to internal molecular E-field, spin substates in a molecule
have slightly different energies. This mechanism leads to a
thermodynamically averaged CP-violating (P,T-odd) magnetic moment
per molecule $\langle \mu^\mathrm{CP} \rangle \sim \mu_B \,  d_e
E_\mathrm{eff} / (k_B T)$, where $E_\mathrm{eff} $ is the large
molecular effective electric field acting on the EDM of the unpaired
electron. $E_\mathrm{eff}$ grows $\propto Z^3$ with the nuclear
charge $Z$ of the heavier molecular constituent \cite{Fla76,San65}
and one would choose to work with heavy radicals. Such molecules as
BaF, YbF, HgF, and HgH belong to this broad category. We found that
mercury hydride (HgH) has parameters most suitable for the proposed
search, see Table~\ref{tab_Wd}. For the HgH molecule
$E_\mathrm{eff}\approx 8 \times 10^{10} \,
\mathrm{V/cm}$~\cite{Koz85} and its ESR spectrum in Ar matrix has
been studied in~\cite{SK02}.

\begin{table}[htb]
\caption{Parameters of several heavy molecules with the ground state
$^2\Sigma_{1/2}$. Molecular dipole moments $D$ were measured in
Refs.~\cite{EKT86,SWH95,NMD89}. Polarization $\langle n_z \rangle$,
the maximal number density $n_\mathrm{max}$, and the accumulation
time $t_\mathrm{acc}$ required to reach S/N=1 for the current limit
on eEDM \eqref{Eq:deTl}. These parameters are calculated with the
help of Eqs.~\eqref{n_z}, \eqref{n_opt}, and~\eqref{Eq:S2Na} for
$E=10$~kV/cm, $T=1$K, and sample volume $0.1$~cm$^3$.}

\label{tab_Wd}

\begin{tabular}{lddddr}
\hline \hline
Molecule
&\multicolumn{1}{c}{$E_\textrm{eff}$\footnote{The effective electric field
for BaF and YbF was calculated in
\cite{KL95,Koz97}. For HgH we rescale results from \cite{Koz85}.}}
&\multicolumn{1}{c}{$D$}
&\multicolumn{1}{c}{$\langle n_z\rangle$}
&\multicolumn{1}{c}{$n_\mathrm{max}$}
&\multicolumn{1}{c}{$t_\mathrm{acc}$}
\\
&\multicolumn{1}{r}{$\left(10^{9}\frac{\rm V}{\rm cm}\right)$}
&\multicolumn{1}{c}{(D)}
&&\multicolumn{1}{c}{$\left(10^{20}\frac{1}{\mathrm{cm}^3}\right)$}
&\multicolumn{1}{c}{(ms)}
\\
\hline
BaF   &     8   &   3.17    &  0.13    &   0.03    &     300    \\
YbF   &    26   &   3.91    &  0.16    &   0.02    &     30     \\
HgH   &    79   &   0.47    &  0.02    &   1.5     &     3      \\
\hline \hline
\end{tabular}
\end{table}

For diatomics, the moment $\langle \mu^\mathrm{CP} \rangle$ is
directed along the molecular axis. For a randomly oriented sample,
however, the net magnetization would vanish. When an external
E-field is applied, it couples to the traditional molecular
electric-dipole moment $D$ and orients the molecules. Taking into
account molecular polarization, the CP-moment can be expressed as
\begin{equation}
    \langle \mu^\mathrm{CP}_\mathrm{mol} \rangle
  \approx \mu_B   \frac{ d_e E_\mathrm{eff}} {k_B T} \times \langle n_z \rangle \, ,
  \label{s_av1}
\end{equation}
where $\langle n_z \rangle$ is the average projection of the
molecular axis onto the E-field (the field is directed along
z-axis). Now the sample acquires a macroscopic magnetization. This
magnetization generates an ultraweak magnetic field $B^\mathrm{CP}$
proportional to eEDM
\begin{equation}
  B^\mathrm{CP} = 4 \pi \gamma \, n \, \langle \mu^\mathrm{CP}_\mathrm{mol} \rangle \, ,
  \label{Eq:BCP}
\end{equation}
where $n$ is the molecular number density and $\gamma$ is a
geometry-dependent factor. For example, near the center of a
disk-shaped sample of radius $R$ and thickness $L$, $\gamma = L/2R$.

Orientation of B-field \eqref{Eq:BCP} is linked to that of the
applied E-field through $\langle n_z \rangle$. Such a link is
forbidden in the traditional electrodynamics. Its very presence is a
manifestation of the parity and time-reversal violation. By
measuring $B^\mathrm{CP}$ one constrains eEDM via
Eqs.~(\ref{Eq:BCP}) and (\ref{s_av1}). It is apparent that
maximizing $n$ is beneficial. However, bringing radicals together is
problematic --- they react chemically. Here is where the matrix
isolation technique~\cite{AndMos89} becomes key. In this
well-established method, the molecules are co-deposited with
rare-gas atoms or other species onto a cold ($T \sim 1 \,
\mathrm{K}$) substrate and become trapped in the matrix. Small
trapped molecules exhibit properties similar to those for free
molecules and a variety of studies, including determination of
hyperfine-structure constants has been carried out.

There is an upper limit on the density of trapped molecules; to
avoid spin alignment in the subsystem of guest molecules one
requires that thermal agitations are stronger than dipole-dipole
interactions between the molecules. We can estimate the maximum
density as:
\begin{align}\label{n_opt}
n_\mathrm{max}\approx\frac{3}{4\pi} \, \frac{k_{B}T}{D^2}.
\end{align}
A particular advantage of HgH is that its dipole moment  is
relatively small, $D=0.47 \, \mathrm{Debye}$~\cite{NMD89} and at
$T=1\, \mathrm{K}$, the density $n_\mathrm{max}\approx 1.5 \times
10^{20} \, \mathrm{cm}^{-3}$.

Estimate \eqref{n_opt} agrees with experimental observations that
1:100 guest to host ratio is possible. According to \cite{KS_priv}
the realistic matrix thickness and area are $L=1$~mm and
$S=1$~~cm$^2$. That corresponds to $\gamma\approx 0.1$ in
\eqref{Eq:BCP}.
Recently developed low density plasma beam source \cite{Ryabov_priv}
produces permanent beam of heavy radicals with intensity $\sim
10^{18}$~mol/sterad/s. Placing 1~cm$^2$ target at 20~cm from the
source, one can accumulate necessary number of radicals, i.e.
$10^{19}$, in 1 hour.

How are the relevant molecular properties modified by the matrix
environment? A free non-rotating molecule may be described by the
electronic wave function  $|\Omega\rangle$, with $\Omega=\pm 1/2$
characterizing projection of spin onto molecular axis. The
time-reversal operation $T$ converts $\Omega$-states into each
other: $|\Omega\rangle \stackrel{T}{\rightarrow} |-\Omega\rangle.$
In the matrix, a molecule can be considered as an individual entity
perturbed by the host atoms. The local symmetry of the perturbing
fields depends on the position of the molecule in the matrix.
Independent of the spatial symmetry the time-reversal symmetry
remains. According to the Kramers' theorem, in the absence of
magnetic fields, all levels of diatomics with half-integer spin
remain two-fold degenerate for any possible electric field.

EDM interaction operates at short distances near the heavier
nucleus. Expanding the electronic wavefunction in partial waves we
notice that contribution to the eEDM signal of total angular momenta
beyond $s_{1/2}$-- and $p_{1/2}$--waves are strongly suppressed
because of the growing centrifugal barrier and properties of the
eEDM  \cite{KL97}. The truncated  wave function has the
$C_{\infty,v}$ symmetry and $\Omega$ still remains a good quantum
number for the degenerate states of matrix-isolated radicals. Within
this approximation, the effective molecular Hamiltonian in the
external field $E$ reads
\begin{equation}\label{H_sr_mtrx}
    H_\mathrm{eff}
    =- \bm{D} \cdot \bm{E^*} + 2 d_e E_\mathrm{eff} \Omega \, ,
\end{equation}
where  $E^*$ is microscopic E-field; for small fields
$E^*=E/\varepsilon$. We used
$H_\mathrm{eff}$ to arrive at Eq.~(\ref{s_av1}).

Using the estimate~(\ref{s_av1}) with the present limit on eEDM~(\ref{Eq:deTl}),
we obtain for the thermally-induced CP-odd
magnetic moment of HgH molecule trapped at $T=1 \, \mathrm{K}$
\begin{equation}\label{Eq:mu_HgH}
    \langle \mu^\mathrm{CP}_\mathrm{mol} (\mathrm{HgH}) \rangle <
     1.4 \times 10^{-12} \langle n_z \rangle \mu_B \, .
\end{equation}
It is instructive to compare this value to the {\em permanent}
molecular CP-violating magnetic moment introduced by us in
Ref.~\cite{DK05}. This moment arises due to a magnetization of the
molecule by its own electric field (irrespective of the
temperature). The largest $\mu^\mathrm{CP}$ for diamagnetic
molecules was found for BiF for which $\mu^\mathrm{CP} < 3 \times
10^{-17} \langle n_z \rangle\mu_B$,  much smaller than the {\em
thermally-induced} CP-odd moment (\ref{Eq:mu_HgH}). Therefore here
we may neglect the {\em permanent} $\mu^\mathrm{CP}$.

An important parameter entering
    $\langle \mu^\mathrm{CP}_\mathrm{mol} \rangle$
is the degree of molecular polarization  $\langle n_z \rangle$ in
the external E-field. Free diatomic molecules can be easily
polarized by the laboratory fields $\sim 10^4$V/cm, but there is a
paucity of data on polarizing matrix-isolated
molecules~\cite{KilSchSch05}. Certainly, the rotational dynamics of
the guest molecule is strongly affected by the matrix cage. The
molecular axis evolves in a complex multi-valley potential, subject
to the symmetry  imposed on the molecules by the matrix cage.
Depending on the barrier height between different spatially oriented
valleys, the guest molecule may either execute hindered rotation or
librations about the valley minima. Ref.~\cite{Khr05} reports
evidence for hindered rotation of HXeBr and Ref.~\cite{Wel90}
suggests that other hydrides can rotate. Note also, that for Ar
matrix the cell size is 4.5\AA, while internuclear distance for HgH
is only 1.7\AA. That gives us a confidence that the HgH radical can
be polarized by the external electric field.

We will distinguish between two limiting cases of molecular polarization:
strong and weak fields. In the former limit $\langle n_{z}\rangle \sim 1$, and
in the latter,
\begin{equation}\label{n_z}
\langle n_{z}\rangle =\frac{1}{Z}\sum_{n_{z}}{n_{z}}\exp\left(
\frac{DE^{\ast}n_{z}}{k_{B}T}\right)
  \approx
\frac{DE^{\ast}}{k_{B}T}\langle n_{z}^{2}\rangle.
\end{equation}
For isotropic orientational distribution, characteristic for the
polycrystalline matrixes, $\langle n_{z}^{2}\rangle = 1/3$, and we
get
\begin{equation}\label{mu_high}
    \langle \mu^\mathrm{CP}_\mathrm{mol} \rangle \approx
    \frac{1}{3} \mu_B \frac{DE^{\ast}}{k_{B}T} \frac{E_\mathrm{eff} d_e}{k_B T} \,.
\end{equation}
The dielectric constant of the rare-gas matrix is close to unity,
but addition of polar molecules results in
\begin{align}\label{eps}
\varepsilon &\approx 1 + 4\pi n \alpha
= 1 +4\pi n \frac{D^2\langle n_{z}^{2}\rangle}{k_{B}T}
\approx 1 +\frac{4\pi}{3} n \frac{D^2}{k_{B}T},
\end{align}
where $\alpha$ is molecular polarizability. For maximum
density~(\ref{n_opt}), $\varepsilon\approx 2$ and $E^{\ast}\approx E/2$.

The parameter differentiating the weak- and the high-field regimes
is the ratio $D E^{\ast}/k_{B}T$. For HgH trapped at $T=1$~K, the
transition occurs at  $E^{\ast} \approx 100 \mathrm{~kV/cm}$. The
breakdown fields for the rare-gas matrices are unknown, we only
notice that for liquid Xe it is 400~kV/cm so that both weak- and
high-field regimes may be possibly realized. The moderate
$E=10$~kV/cm field corresponds to $\langle n_{z}\rangle \approx
0.02$.

Finally, we proceed to evaluating the sensitivity of the proposed
eEDM search. There are two crucial criteria to consider: weakest
measurable B-field and signal-to-noise ratio. Presently, the most
sensitive measurement of magnetic fields has been carried out by the
Princeton group (see \cite{KomKorAll03} and references therein).
This group has reached the sensitivity level of $5.4 \times
10^{-12}\, \mathrm{G}/\sqrt{\mathrm{Hz}}$. A projected experimental
sensitivity of $3 \times 10^{-15} \, \mathrm{G}/\sqrt{\mathrm{Hz}}$
is published in \cite{Lam02}. We find that for $\langle n_{z}\rangle
\sim 1$ and for $\gamma=0.1$ the present eEDM limit may be recovered
within integration time of $t=5\, \mathrm{s}$ for the demonstrated
sensitivity and within $10^{-6} \, \mathrm{s}$ for the projected
sensitivity. Alternatively, during a week-long measurement, the
present eEDM limit may be improved by $3 \times 10^2$ for the
demonstrated and by $6 \times 10^5$ for the projected B-field
sensitivity. These values are reduced by a factor of 50 for a
moderate 10 kV/cm polarizing field.

In addition to limitations imposed by the weakest measurable B-field
one must also consider signal-to-noise ratio \cite{BLS06}. As we
pointed out above, the thermally-induced $\langle
\mu^\mathrm{CP}_\mathrm{mol} \rangle$ of radicals is many orders
larger than permanent $\mu^\mathrm{CP}_\mathrm{mol}$ of diamagnetic
molecules discussed in~\cite{DK05}. The magnetic noise from
paramagnetic radicals is also much higher as they have traditional
magnetic moments associated with unpaired electron spin,
\begin{equation}\label{mu_t}
\langle \mu_\mathrm{mol} \rangle = 2\mu_B \Omega\langle
n_{z}\rangle.
\end{equation}
These moments lead to random magnetization of the sample and
generate a fluctuating B-field. Unlike $B^\mathrm{CP}$, this field
is not correlated with the direction of the external E-field and it
is the main source of the noise. In our case, the signal-to-noise
ratio is
\begin{equation} \label{Eq:S2N}
S/N = 3 \frac{\langle \mu^\mathrm{CP}_\mathrm{mol} \rangle}{\mu_B}
 \sqrt{\mathcal{N}\frac{t}{\tau}} \, ,
\end{equation}
where $\mathcal{N}$ is the number of molecules, $t$ is the
observation time, and $\tau$ is the correlation time for the random
thermal magnetization. Factor 3 at the right hand side appears
because of the averaging of the magnetic moment \eqref{mu_t} over
orientations of the molecular axis $\bm{n}$.

For a strong spin-rotation coupling, as in the case of HgH, $\tau$ is
determined by interaction of molecular axis with environment. One of
such mechanisms is the dipolar interaction between guest radicals,
so that $\tau \sim \hbar/(D^2 n) = 4 \pi \hbar/(3 k_B T)$ for the
optimal density~\eqref{n_opt}.
For the weak-field limit
\eqref{mu_high} we get the final expression for S/N:
\begin{equation} \label{Eq:S2Na}
S/N = \frac{3}{8\pi} \frac{E E_\mathrm{eff} d_e}{k_B T}
\sqrt{Vt/\hbar}\, ,
\end{equation}
where $V$ is the sample volume. This equation is used in
Table~\ref{tab_Wd} to estimate accumulation time needed to reproduce
the current limit \eqref{Eq:deTl}. For HgH molecule we find that for
a volume of $0.1 \, \mathrm{cm}^3$ and strong polarizing field, the
present eEDM limit may be recovered within $t=10^{-6}\, \mathrm{s}$
(3~ms for the field 10~kV/cm). By integrating the signal for one
week, the present eEDM limit may be improved by a factor of $2
\times 10^6$. Note that these estimates are close to the estimates
based on the projected sensitivity to the weak magnetic
fields~\cite{Lam02}.


To summarize, our proposed eEDM search combines advantages of the
strong intermolecular field with a high attainable number density of
molecules embedded in a matrix of rare-gas atoms. We argue that our
proposal has a potential of improving the present eEDM limit by
several orders of magnitude. That will allow constraining the ``new
physics'' beyond the Standard Model at an important new level and,
in particular,
testing predictions of competing  SUSY models.

{\em Acknowledgments.} We would like to thank  D.~Budker, T.~Isaev,
L.~Knight, S.~Lamoreaux, S.~Porsev, M.~Romalis, V.~Ryabov,
I.~Savukov, R.~Sheridan, O.~Sushkov, and I.~Tupitsyn for valuable
comments and discussions. This work is supported by the Russian
Foundation for Basic Research, grant No.~05-02-16914, by NSF and
NIST.


\end{document}